# Current-polarity dependent manipulation of antiferromagnetic domains

P. Wadley[1*†], S. Reimers[1,2†], M. J. Grzybowski[3], C. Andrews[1], M. Wang[1], J.S. Chauhan[1], B.L. Gallagher[1], R.P. Campion[1], K.W. Edmonds[1], S.S. Dhesi[4], F. Maccherozzi[4], V. Novak[5], J. Wunderlich[6,5], T. Jungwirth[5,1]

1. School of Physics and Astronomy, University of Nottingham, Nottingham NG7 2RD, United Kingdom
2. I. Physikalisches Institut, Friedrich-Hund-Platz 1, 37077 Göttingen, Germany
3. Institute of Physics, Polish Academy of Sciences, Aleja Lotnikow 32/46, PL-02668 Warsaw, Poland
4. Diamond Light Source, Chilton, Didcot, Oxfordshire, OX11 0DE, United Kingdom
5. Institute of Physics, Academy of Sciences of the Czech Republic, Cukrovarnicka 10, 162 00 Praha 6, Czech Republic
6. Hitachi Cambridge Laboratory, J. J. Thomson Avenue, Cambridge CB3 0HE, United Kingdom

**Antiferromagnets have a number of favourable properties as active elements in spintronic devices, including ultra-fast dynamics, zero stray fields and insensitivity to external magnetic fields[1]. Tetragonal CuMnAs is a testbed system in which the antiferromagnetic order parameter can be switched reversibly at ambient conditions using electrical currents[2]. In previous experiments, orthogonal in-plane current pulses were used to induce 90° rotations of antiferromagnetic domains and demonstrate the operation of all-electrical memory bits in a multi-terminal geometry[3]. Here, we demonstrate that antiferromagnetic domain walls can be manipulated to realize stable and reproducible domain changes using only two electrical contacts. This is achieved by using the polarity of the current to switch the sign of the current-induced effective field acting on the antiferromagnetic sublattices. The resulting reversible domain and domain wall reconfigurations are imaged using x-ray magnetic linear dichroism microscopy, and can also be detected electrically. The switching by domain wall motion can occur at much lower current densities than those needed for coherent domain switching.**

*email: Peter.Wadley@nottingham.ac.uk †These authors contributed equally to this work

Magnetization reversal in a ferromagnet typically proceeds by the nucleation and propagation of magnetic domain walls (DWs). The magnetic field where reversal occurs is determined by the pinning strength of local defects, rather than the (usually much larger) magnetic anisotropy energy barrier. In a ferromagnetic wire, DWs can be manipulated electrically by current-induced spin transfer torque or spin-orbit torque[4].

In an antiferromagnetic material, neighbouring atomic magnetic moments are aligned to give zero net magnetization, resulting in an insensitivity to magnetic fields of up to tens of teslas. Switching an antiferromagnet's staggered magnetization can be realized by passing uniform electrical currents through the antiferromagnet if certain symmetry conditions are met. Firstly, if a site in the lattice has a local inversion asymmetry, a uniform current induces a local non-equilibrium spin polarization at that site[5]. Secondly, if the antiferromagnetic sublattices are space-inversion partners, the current-induced spin polarization generates an effective magnetic field ($H_{eff}$) whose sign alternates with each sublattice. This staggered field can induce a Néel-order spin-orbit torque (NSOT)[6], $m_i \times H_{eff}$ where $m_i$ is the sublattice magnetization, which can efficiently reorient the antiferromagnetic moments. These conditions are fulfilled in, among other systems, the room-temperature collinear antiferromagnets CuMnAs[7] and Mn$_2$Au[5], where current-induced switching was recently demonstrated[2,3,8,9,10].

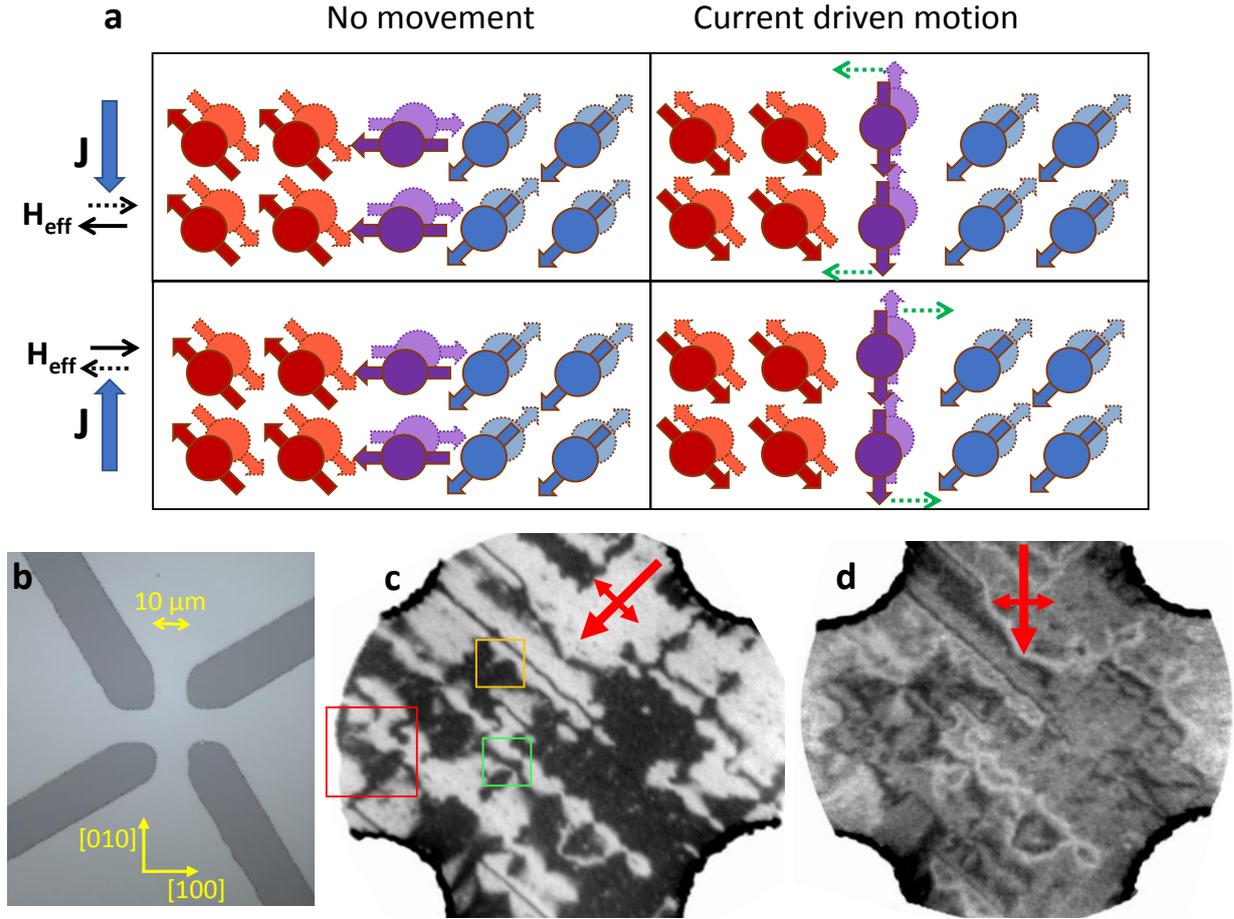

**Figure 1. Antiferromagnetic DW motion concept and device structure. a**, Four possible configurations, relative to the applied current direction, of an antiferromagnetic domain wall in the *ab* plane of tetragonal CuMnAs. Spins on opposite sides of the domain wall are shown in red and blue, respectively, with the spins in the domain wall shown in purple. The current direction (**J**) is shown by the thick blue arrows on the left side of the figure, while the Néel-order current-induced effective fields (**H**$_{eff}$) acting on each sublattice are shown by the solid and dotted black arrows respectively. The green dotted arrows show the direction of propagation of the domain wall. **b**, Optical micrograph of the CuMnAs cross. The darker patches are where the CuMnAs has been removed by chemical etching down to the GaP substrate. **c**, Antiferromagnetic domain structure observed for x-ray polarization **E** // [1-10] crystal axis. **d**, Antiferromagnetic domain structure observed for **E** // [100] crystal axis. The single- and double-headed arrows in **c** and **d** indicate the in-plane projection of the x-ray propagation vector and the x-ray polarization vector, respectively. The light (dark) contrast corresponds to antiferromagnetic domains with staggered magnetization oriented perpendicular (parallel) to the x-ray polarization for **E** // [1-10]. This ordering is reversed for **E** // [100] (see Methods). The red, orange and green squares in **c** correspond to the regions shown in Fig. 2, Fig. 3b and Fig. 3d respectively.

Here, we demonstrate reversible switching in an epitaxial CuMnAs film, controlled by the *polarity* of an applied current pulse ($I_{pulse}$). In previous studies of antiferromagnetic memory devices, the reported electrical read-out signal was polarity-independent; reconfigurations of the antiferromagnetic moments were only achieved by applying current pulses in orthogonal directions[2,3,8]. On a microscopic level, however, the polarity of the writing current can play an important role as it determines the DW propagation direction[11,12]. This is illustrated in Fig. 1a, for current applied at 45° to the biaxial easy magnetic axes of the CuMnAs crystal (*i.e. I$_{pulse}$* along the [010] and [0-10] directions). The out-of-plane *c*-axis is energetically unfavourable in tetragonal CuMnAs[13], so the spins align in the plane of the film throughout the DW. The current-induced field **H**$_{eff}$ is orthogonal to the current direction, and in opposite directions for each sublattice. As shown in Fig. 1a, antiferromagnetic DW motion due to NSOT can be expected when the spins in the DW are aligned perpendicular to **H**$_{eff}$. Then, there is a torque acting on the wall whose sign depends on the polarity of *I$_{pulse}$*. Moreover, the strong intersublattice exchange coupling in an antiferromagnet is predicted to result in much faster DW motion compared to ferromagnets, where the velocity is limited by Walker breakdown[11,12,14,15].

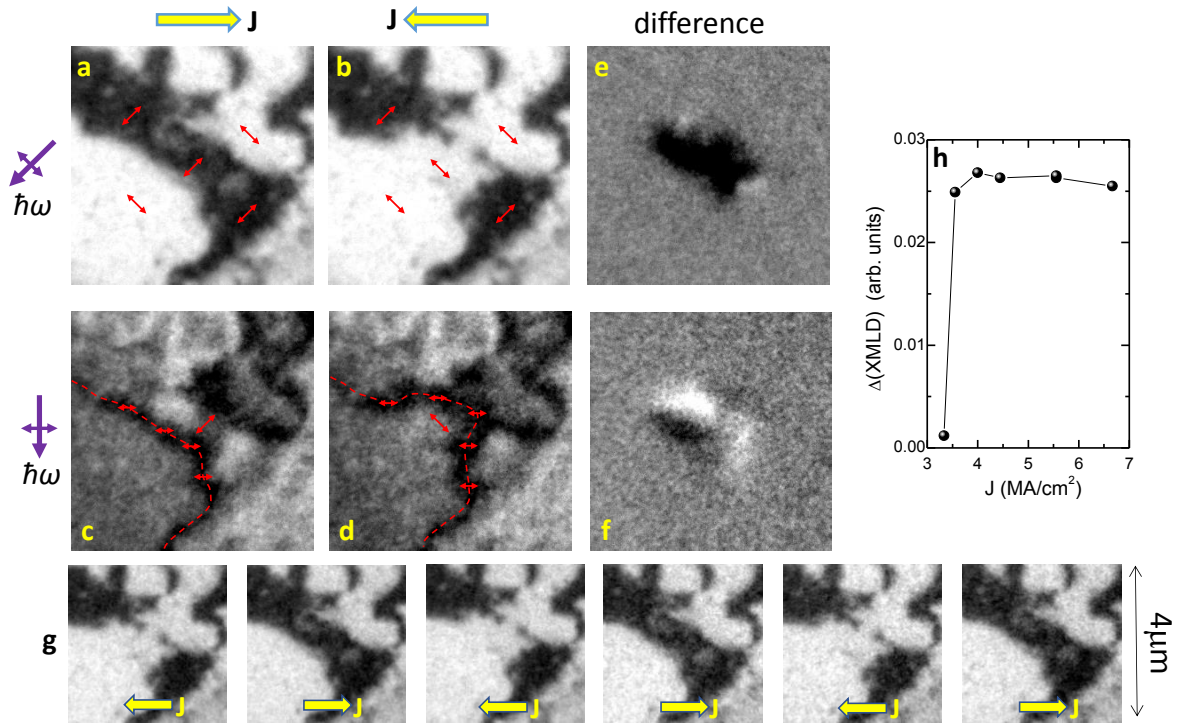

**Figure 2. Antiferromagnetic domain switching by current-induced domain wall motion. a-d**, 4mm x 4mm XPEEM images of an antiferromagnetic domain structure located along one of the arms of the CuMnAs cross, obtained with **E** //[1-10] axis (**a,b**) and **E** // [100] axis (**c,d**), after applying current pulses in the directions shown at the top of the figure. The red double-headed arrows indicate the local orientation of the staggered magnetization, as inferred from the XPEEM images. The dashed red line in **c** and **d** marks the domain wall which is affected by the current pulse. **e**, Difference between the images shown in panels **a** and **b**. **f**, Difference between the images shown in panels **c** and **d**. **g**, Same as for **a**,**b** but for several reversals of the current pulse polarity. **h**, Dependence of the average change in signal on the current density of the pulse, for a fixed pulse duration of 2.5ms.

Figures 1b-d show an optical micrograph and antiferromagnetic domain images for a 10µm cross structure fabricated from a 45nm thick CuMnAs film. The domain images were obtained using x-ray photoemission electron microscopy (XPEEM), with contrast due to x-ray magnetic linear dichroism (XMLD)[8]. The XMLD signal varies as $A\cos^2\phi$, where $\phi$ is the angle between the magnetic moment and the x-ray polarization vector (**E**). The prefactor *A* depends on the orientation of **E** with respect to the crystal axes[16,17] (see Methods). The large black-white contrast observed for **E**||[1-10] (Fig. 1c) indicates a biaxial magnetic anisotropy, with the antiferromagnetic moments aligned predominantly along the [110] and [1-10] axes. For **E**||[100] (Fig. 1d), the black-white contrast is predominantly at the boundaries of the domains seen in Fig. 1c.

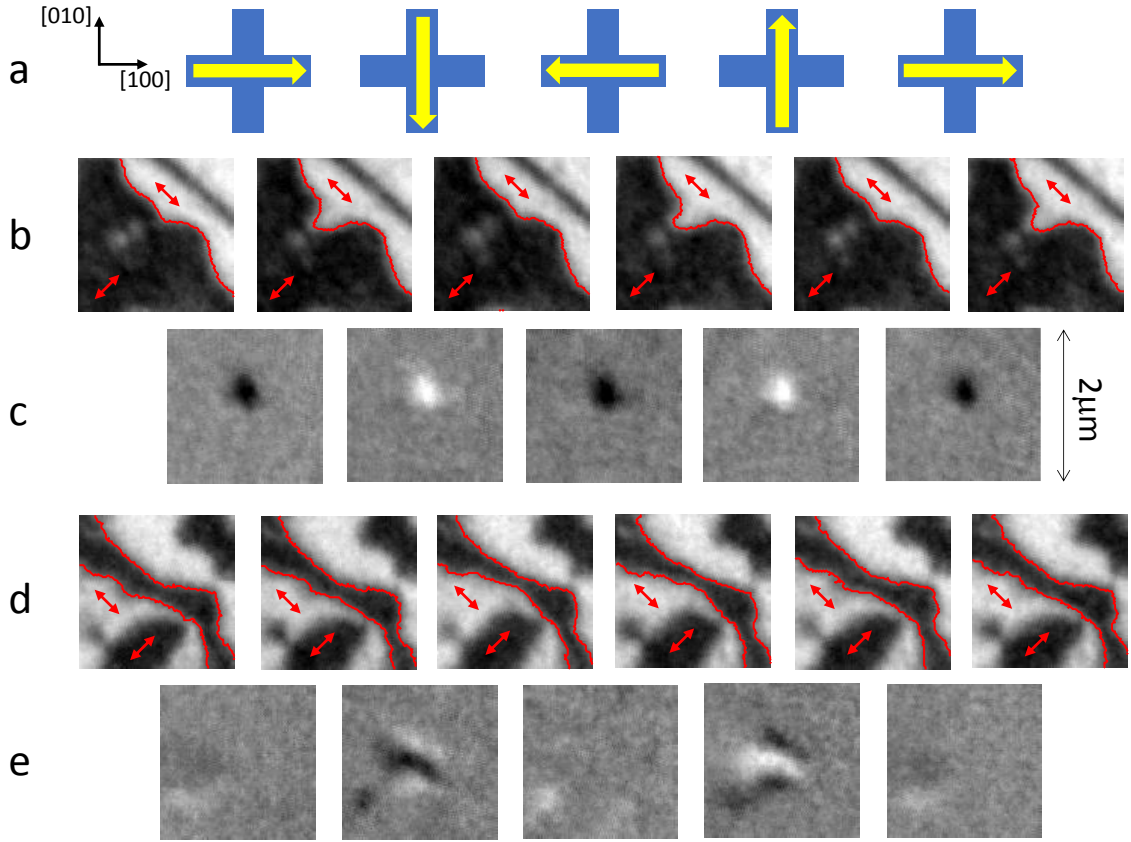

**Figure 3. Dependence on the current pulse direction. a**, Illustration of the current pulse sequence. **b,** Expanded views of the 2mm x 2mm region marked by the orange square in Fig. 1c, after applied current pulses in the directions shown schematically in **a**. **c**, Difference between the two magnetic domain images shown directly above in **b**. **d,** Expanded views of the 2mm x 2mm region marked by the green square in Fig. 1c, after applied current pulses in the directions shown schematically in **a**. **e**, Difference between the two magnetic domain images shown directly above in **d**. The red lines in mark the 50% contour between the black-white regions. The double-headed arrows represent the staggered magnetization orientation within each domain.

Fig. 2 shows polarity-dependent switching of an antiferromagnetic domain within the 4μm x 4μm region marked by the red square in Fig. 1c. Electrical current pulses were applied along the [100] oriented arm of the device, with alternate polarity. Comparison of Fig. 2a and Fig. 2b shows the switching of the centre of the image from dark to light, corresponding to a rotation of the staggered magnetization from the [110] to the [1-10] axis. Fig. 2c,d show the same region viewed with **E**||[100], where sensitivity to domain walls is expected, and here it can be seen that an approximately 100nm wide DW has moved from one side of the switched region to the other. The differences between the images obtained after opposite polarity current pulses, shown in Fig. 2e,f, highlight that the current-induced switching is localized in a ≈2μm x 1μm region in the centre of the image. Fig. 2g shows the reproducibility of the switching over several polarity reversals of $I_{pulse}$. The net difference between the domain images, averaged across the whole 4μm x 4μm square, shows a sharp onset at a current density of ≈3.5 MA/cm$^2$, with no further changes up to the highest currents applied (Fig. 2h).

Next, we show that both the originally reported polarity-independent orthogonal switching[2,3,8] and the newly observed polarity-dependent switching can exist in different regions of the same device. This is achieved by examining the effect of applying $I_{pulse}$ in different directions through the centre of the cross. Current pulses of amplitude 6MA/cm$^2$ were applied sequentially in the directions illustrated in Fig. 3a. We focus on two 2μm square regions, marked by the orange and green squares in Fig. 1c, which show pronounced effects. The region displayed in Fig. 3b,c shows a clear switching with each current pulse with orthogonal dependence: the centre of the image shifts from black to white when $I_{pulse}$ is along the [100] axis, and white to black for $I_{pulse}$ along the [010] axis, independent of the polarity. In Fig. 3d,e, negligible changes are observed for $I_{pulse}$ along the [100] axis, while a shift in the DW running through the centre of the image is observed for $I_{pulse}$ along the [010] axis. The polarity-dependence can be clearly seen in Fig. 3e, which shows the difference between images obtained before and after each current pulse: current pulses along [010] and [0-10] result in difference images of opposite contrast.

We now demonstrate that the polarity-dependent switching of antiferromagnetic domains can result in a measurable electrical signal. Fig. 4a shows the 8-arm CuMnAs device used for the electrical study. The longitudinal and transverse resistances, $R_L = V_L / I_{probe}$ and $R_T = V_T / I_{probe}$, of the device were measured using the geometry illustrated in Fig. 4a, following the application of $I_{pulse}$ of alternating polarity. As shown in Fig. 4b, the measured $R_T$ depends on whether the current pulse was applied with positive or negative polarity, *i.e.* there is a memory effect. The current-induced change in $R_T$ shows an onset for current pulses of around 25mA and reaches a plateau at around 30mA (Fig. 4c), which corresponds to a current density of around 3MA/cm$^2$ in the centre of the device. In contrast, no measurable change of $R_L$ was observed.

The observed change in $R_T$ is qualitatively consistent with the domain switching by DW propagation discussed previously. The current pulses are applied at 45$^o$ to the biaxial easy axes of the antiferromagnetic film which, as illustrated in Fig. 1a and shown directly in Fig. 2a,b, can lead to a rotation of the staggered magnetization between +45$^o$ to -45$^o$ with respect to the current direction. This in turn can produce a transverse anisotropic magnetoresistance (AMR) in the probe geometry, which varies as sin2$\theta$, where $\theta$ is the angle between the staggered magnetization and the probe current direction. The small size of the transverse AMR shown in Fig. 4c is consistent with the domain switching only occurring in localized regions within the central cross. We have observed similar behaviour in several samples patterned from the same wafer, but the onset current amplitude and the size of the read-out signal vary considerably. This supports a picture where the local pinning of a DW dictates the threshold current and the domain size can vary. The lowest threshold current density was 0.5 MA/cm$^2$ for one of the devices tested. In contrast, switching using orthogonal current pulses required current densities above 4 MA/cm$^2$ for the same devices. The close correlation between XPEEM images and transverse AMR for orthogonal current pulse switching has been shown previously[2,8].

Our results show directly the switching of antiferromagnetic domains, in a way that depends on the polarity of the current pulse. This offers not only a straightforward two-terminal device geometry but potentially much lower writing current compared to previous studies of antiferromagnetic memory devices. In principle, the mechanism we ascribe to polarity-dependent switching could apply both to switching via DW motion as well as rotation of an entire domain. However, the latter should require substantially higher spin-orbit effective fields to overcome the hard axis anisotropy energy barrier. The displacement of the domain wall shown in Fig. 2c,d, where the spins in the DW are collinear with the current direction, strongly points towards DW motion as the origin of the observed domain switching. This provides a route to exploring the novel physics of antiferromagnetic domain wall motion[11,12,14,15].

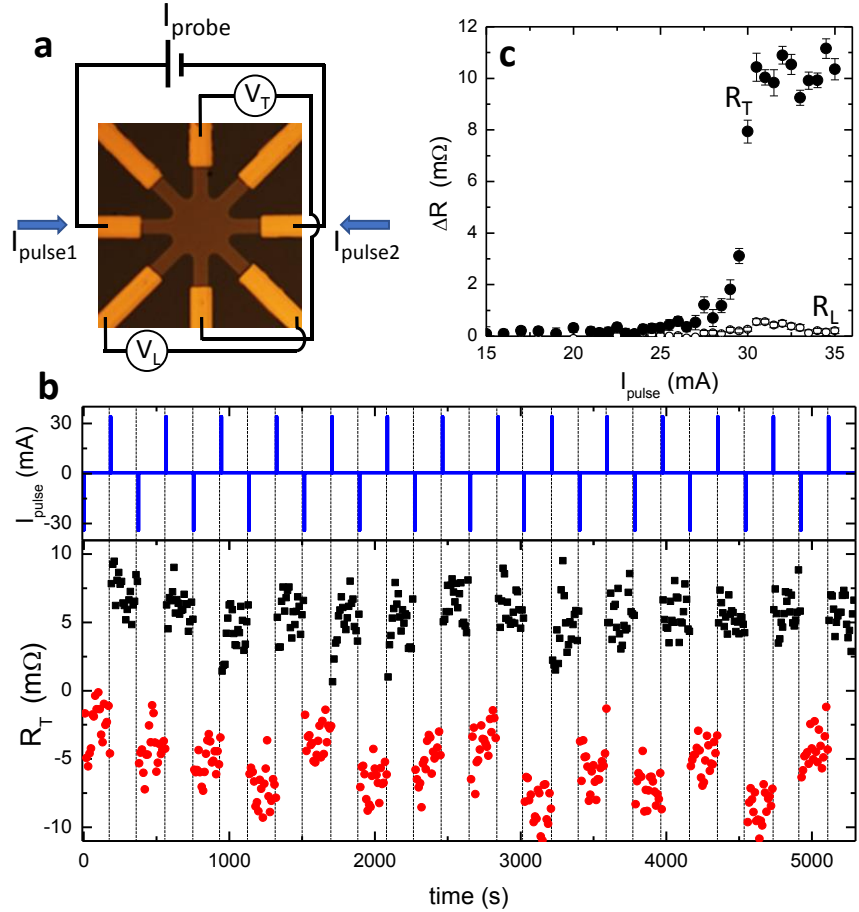

**Figure 4. Electrical detection of current-induced switching. a,** Optical micrograph of the CuMnAs device, consisting of 8 arms of width 10mm, and schematic of the experiment geometry. After applying current pulses $I_{pulse1}$, $I_{pulse2}$ of duration 50ms and varying amplitude, the longitudinal and transverse voltages $V_L$ and $V_T$ are recorded for a probe current $I_{probe}$ of 500mA. **b,** Time series showing the applied current pulses (top) and the transverse resistance $R_T = V_T/I_{probe}$ measured following current pulses of +34mA (black) and -34mA (red). A constant offset has been subtracted from $R_T$. The vertical lines are guides to the eye. **c,** Measured change in the transverse resistance (filled circles) and longitudinal resistance (open circles) versus the current pulse amplitude.

The observed domain structure, and the fact that current-induced switching occurs only in localized regions of the structure, suggests that the pinning of domain walls plays important role, for example due to local defects and strains. The underlying pinning mechanisms may be significantly different than in ferromagnetic DWs; for example, stray-field effects are much weaker, while magnetoelastic and strain effects may be much more important. Both strain and the magnetoelastic constant should be readily influenced by varying the substrate lattice parameters and/or chemical composition. Elucidation of these mechanisms will provide a route towards designing ultrafast, high efficiency AF DW-based spintronics devices.

## Methods

**Sample preparation.** The 45nm CuMnAs film was grown on a GaP buffer layer on a GaP(001) substrate at 260°C, and capped with a 2nm Al film to prevent surface oxidation. In-situ reflection high energy electron diffraction measurements confirmed the tetragonal structure of the layer, with the epitaxial relationship CuMnAs(001)[100]||GaP(001)[110][7]. The crystallographic axes described in the main text correspond to those of the CuMnAs film. The device structures were fabricated by optical lithography and wet etching, with electrical contacts made by wire-bonding to Cr/Au pads.

**XPEEM imaging.** The XPEEM measurements were performed at room temperature on beamline I06 at Diamond Light Source, using linearly polarized x-rays. Magnetic contrast with approximately 30nm spatial resolution was obtained from the difference in the absorption signal measured by XPEEM at the peak and the minimum of the Mn $L_3$ XMLD spectrum. The crystalline dependence of the XMLD was determined previously using an exchange-coupled Fe/CuMnAs bilayer, where it was shown that the XMLD spectrum has a similar shape but opposite sign for **E**||[100] and **E**||[110] [17]. The electrical current pulses of duration 2.5ms and varying amplitude were applied *in-situ*.

**Electrical measurements.** All the electrical measurements were performed at room temperature on 8-arm devices, using the geometry illustrated in Fig. 4a. The width of each arm is 10μm. Current pulses of duration 50ms and varying amplitude were applied. Probing was performed using dc currents of amplitude 0.5mA (0.05 MA/cm$^2$). The resistivity of the devices was around $2\times10^{-4}$ Ωcm at room temperature.

## Acknowledgements

We thank the Diamond Light Source for the allocation of beam time under Proposal No. SI16376-1. We acknowledge support from Engineering and Physical Sciences Research Council grant EP/P019749/1, National Science Centre, Poland (grant 2016/21/N/ST3/03380), the Ministry of Education of the Czech Republic Grants No. LM2015087 and No. LNSM-LNSpin, the Czech National Science Foundation Grant No. 14-37427, the EU FET Open RIA Grant No. 766566, and the ERC Synergy Grant No. 610115. P.W. acknowledges support from the Royal Society through a University Research Fellowship.

## Author contributions

P.W. , K.W.E., B.L.G., J.W. and T.J. were responsible for the experimental concept and design. K.W.E. and P.W provided experimental coordination of the project. R.P.C. and V.N. performed the material growth. J.S.C., P.W. and C.A. provided device design and photolithography of devices. S.R., M.J.G., K.W.E, P.W., F.M., S.S.D. performed XMLD-PEEM measurements and analysis of results. S.R. performed electrical transport measurements and analysis. M.W. supplied magnetometry measurements of the materials. All authors contributed to the interpretation of the results and writing of the manuscript.